\documentclass[prl,preprint,epsfig,superscriptaddress,rotate,noshowpacs]{revtex4}

\usepackage[version=3]{mhchem} 
\usepackage[T1]{fontenc}       
\usepackage{epsfig} 
\usepackage{graphicx} 
\usepackage{amssymb, amstext, amsmath} 
\usepackage{bbold} 
\usepackage{hyperref} 
\usepackage{color}
\usepackage{verbatim}
\usepackage{mathrsfs}
\usepackage[normalem]{ulem}

\begin{document} 

\title{A Model of Charge Transfer Excitons: Diffusion, Spin Dynamics, and Magnetic Field Effects}
\author{Chee Kong Lee}
\affiliation{Department of Chemistry, Massachusetts
Institute of Technology, Cambridge, Massachusetts 02139, USA}
\author{Liang Shi} 
\affiliation{Department of Chemistry, Massachusetts
Institute of Technology, Cambridge, Massachusetts 02139, USA}
\author{Adam P. Willard} \email{awillard@mit.edu}
\affiliation{Department of Chemistry, Massachusetts
Institute of Technology, Cambridge, Massachusetts 02139, USA}

\begin{abstract} 
In this letter we explore how the microscopic dynamics of charge transfer (CT) excitons are influenced by the presence of an external magnetic field in disordered molecular semiconductors. This influence is driven by the dynamic interplay between the spin and spatial degrees of freedom of the electron-hole pair. 
To account for this interplay we have developed a numerical framework that combines a traditional model of quantum spin dynamics with a stochastic coarse-grained model of charge transport. 
This combination provides a general and efficient methodology for simulating the effects of magnetic field on CT state dynamics, therefore providing a basis for revealing the microscopic origin of experimentally observed magnetic field effects. 
We demonstrate that simulations carried out on our model are capable of reproducing experimental results as well as generating theoretical predictions related to the efficiency of organic electronic materials. 
\end{abstract}

 \maketitle


Charge transfer~(CT) states play a fundamental role in mediating interconversion between bound electronic excitations and free charge carriers in organic electronic materials.
For processes that require this interconversion, such as electroluminescence in organic light emitting diodes (OLEDs) and photocurrent generation in organic photovoltaics (OPVs), low-energy (thermalized) CT states are often implicated as a precursor to efficiency loss pathways
\cite{Segal2007, Difley2008,XYZhu2009, Deibel2010, bredas2009, vandewal2009,Veldman2009,Clarke2010,Bakulin2012,Jailaubekov2013,gelinas2014,Zhu2015}.
Despite this, much remains to be understood about the properties of CT states and how they contribute to various energy loss mechanisms. 
Due to their short lifetime and low optical activity, attempts to interrogate CT states directly have brought limited success.
Notably, however, recent experiments that probe CT states indirectly via their response to an applied magnetic field have demonstrated the potential to reveal new information about this elusive class of excited states\cite{Hu2007, Wang2008,Hu2009, Kersten2011,Kersten2011a,Chang2015a, Deotare2015, Wang2016,Hontz2015}.
Unfortunately, extracting this information is challenging because it is encoded by a complex interplay of electronic and nuclear spin dynamics\cite{Steiner1989,Frankevich1992,Hu2009}.
This interplay is further complicated when the dynamics of the electron-hole spin state (or the specific experimental observable) is coupled to a source of fluctuating microscopic disorder such as charge transport or molecular conformational dynamics\cite{Hontz2015}. 
In this letter we focus on disentangling this interplay.

\begin{figure}[t]   \center
    \includegraphics[width=6.5in]{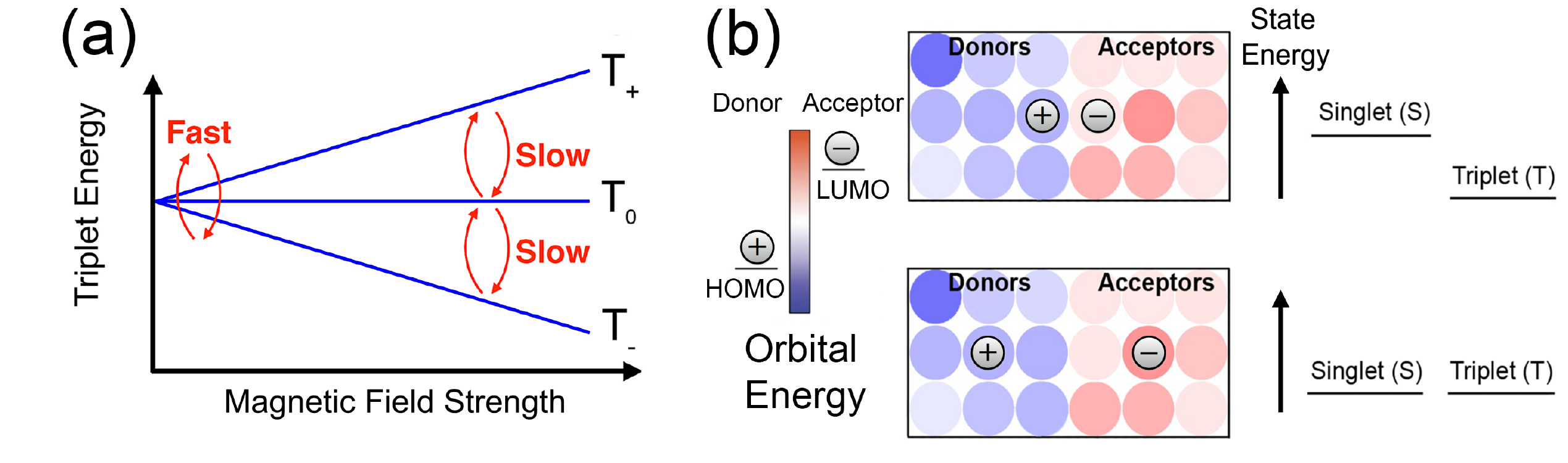}
    \caption{
    (a) The relative energy of the three, otherwise degenerate, triplet levels are split (Zeeman splitting) by a magnetic field.
    Consequently, as indicated with red arrows, the timescale for spin mixing dynamics can be varied with an external magnetic field. 
    (b) A schematic depiction of the model system in which donor and acceptor molecules (represented as blue- and red-shaded circles respectively) reside on opposite halves of an ordered lattice. 
    The electron (-) and hole (+) occupy individual molecules whose orbital energies vary as indicated by shading. 
    The relative energy of singlet and triplet states is determined by the exchange splitting, which decays rapidly with electron-hole separation.
    When the electron and hole occupy neighboring sites (top) this exchange splitting is typically larger than thermal energies.
    When the electron and hole are separated by one or more molecules (bottom) the exchange splitting is negligible resulting in degenerate singlet and triplet energy levels.
    }
    \label{fig:schematic}
\end{figure}

The dependence of an experimental observable on an applied magnetic field is generically referred to as the magnetic field effect (MFE).
For CT-mediated processes, MFEs require that the observed physical property depends either directly or indirectly on the spin state of the electron-hole pair. 
For instance, spin selection rules for radiative electron-hole recombination can give rise to a magnetic field-dependent electroluminescence yield \cite{Desai2007,Macia2014,Crooker2014,Wang2016}.
To understand specifically how CT state properties are influenced by the presence of a magnetic field it is natural to describe the spin state of the electron-hole pair in a standard basis of singlet and triplet states. 
If the electron and hole positions are static then MFEs emerge when the Zeeman splitting of the triplet energy levels becomes comparable to or larger than interactions that govern population transfer between the three triplet spin states (i.e., $T_-$, $T_0$, and $T_+$)~\cite{Steiner1989,Hontz2015}.
Under typical experimental conditions (i.e., applied field strengths $\sim1$T) the magnitude of the Zeeman splitting is much smaller than the thermal energy (i.e., $\Delta E_\mathrm{Zeeman} \ll k_\mathrm{B} T$) and thus it has negligible effect on equilibrium properties. 
The net result, as illustrated in Fig.~\ref{fig:schematic}a, is that the timescale for spin mixing dynamics is slowed in the presence of a magnetic field. 

The microscopic origin of MFEs becomes more complicated if the electron and hole positions are dynamic.
This is because variations in electron-hole separation can drive fluctuations in the value of the exchange coupling that determines the energy difference between the singlet and triplet states.
This coupling can be large compared to thermal energies but decays exponentially with electron-hole separation. 
Even subtle changes in CT state configuration can result in significant variations in the equilibrium singlet-triplet ratio. 
The ability of the CT spin state to respond to these time-dependent variations is mediated by the timescale for spin-mixing dynamics, which as described above, can be tuned by the application of an external magnetic field.
It is this competition of timescales, between spin and spatial dynamics, that ultimately determines the magnitude of the observed MFEs. 
Perhaps more importantly though, is that the MFEs encode information that can be used to characterize the microscopic dynamics of the electron-hole pair. 

In this letter we address the challenge of predicting and interpreting the results of magnetic field sensitive experiments by utilizing numerical simulation.
We present a model for CT state dynamics that incorporates magnetic field dependent spin dynamics into an efficient coarse-grained description of electron-hole transport.
We demonstrate that this model is capable of reproducing experimental results and revealing fundamental aspects of CT state dynamics that are otherwise inaccessible to current experimental techniques.
First, however, we describe the general theoretical framework that underlies our model for simulating CT state dynamics in disordered molecular systems. 

In our model a CT state is described as an oppositely charged pair of spin-1/2 particles (i.e., an electron and hole), each localized on separate molecules. 
We utilize a coarse-grained description of a molecular semiconductor in which individual molecules are represented as discrete sites.
Each molecular site is characterized by a position, $\vec{r}_i$, HOMO energy, $E_i^{(\mathrm{HOMO})}$, LUMO energy, $E_i^{(\mathrm{LUMO})}$, and a hyperfine magnetic field, $\vec{B}_i^{(\mathrm{hf})}$, which arises from the interaction of the electronic magnetic moment with the nuclear magnetic moment.
CT state properties are determined by combining these parameters, for the electron- and hole-occupied sites, with a description of the electron-hole spin state, which we represent in terms of a two-spin quantum density matrix, $\rho$. 
The energy of a CT state configuration in which the electron occupies site $i$ and the hole occupies site $j$ is given by
\begin{equation}
E_{ij}^{(\mathrm{CT})}(\rho) = E_{ij}^{(\mathrm{E})} + E_{ij}^{(\mathrm{S})}(\rho),
\end{equation} 
where $E_{ij}^{(\mathrm{E})}$  is the electronic energy and $E_{ij}^{(\mathrm{S})}(\rho)$ is the spin energy. 
The electronic energy, which depends on the spatial configuration of the electron-hole pair, is given by 
\begin{equation}
E_{ij}^{(\mathrm{E})}=E_i^{(\mathrm{LUMO})} - E_j^{(\mathrm{HOMO})} - \frac{e^2}{4 \pi \epsilon \vert \vec{r}_i - \vec{r}_j \vert},
\label{eq:cg}
\end{equation}
where $e$ is the elementary unit of charge, and $\epsilon$ is the dielectric constant.
In this expression the first two terms represent the vertical excitation energy (i.e., the HOMO-LUMO gap) of the given CT state and the final term describes the electrostatic electron-hole attraction\cite{Dreuw2004}.
The spin energy is given by 
$E_{ij}^{(\mathrm{S})} = \mathrm{Tr}[H^\mathrm{(S)}_{ij} \rho]$, 
where $H^{(\mathrm{S})}_{ij}$ is the spin Hamiltonian,
\begin{equation}
H_{ij}^{(\mathrm{S})}= g\mu_b\Big[(\vec{S}_e + \vec{S}_h)\cdot \vec{B}^{(\mathrm{app})} +  \vec{S}_e\cdot \vec{B}^{(\mathrm{hf})}_i  +  \vec{S}_h\cdot \vec{B}^{(\mathrm{hf})}_j \Big]  - J(\vert \vec{r}_i - \vec{r}_j \vert) \vec{S}_e \cdot \vec{S}_h, 
\label{eq:hamiltonian}
\end{equation}
where $\mu_b$ is the Bohr magneton, and $g$ is the g-factor for the magnetic moment, $\vec{S}_e$ and $\vec{S}_h$ are the spin operators for the electron and hole respectively.
The terms in the square brackets describe the interaction of the electron and hole spins with the applied magnetic field and the local hyperfine field, denoted as $\vec{B}^{(\mathrm{app})}$ and $\vec{B}^{(\mathrm{hf})}_i$ respectively. 
To model the hyperfine interaction with the nuclear spins we adopt the semiclassical approach of Schulten and Wolynes, in which hyperfine interactions are approximated to be static and site dependent, with $\vec{B}^{(\mathrm{hf})}_i$ drawn randomly from the three-dimensional Gaussian distribution\cite{Schulten1978,Harmon2012a}.
The final term in Eq.~\ref{eq:hamiltonian} describes the exchange interaction between the electron and hole spins, where $J(r)$ is the exchange coupling, which depends on the electron-hole separation, $r=\vert \vec{r}_i - \vec{r}_j \vert$.

The time evolution of our model is separated into a spatial part, which describes the dynamics of electron and hole positions, and a spin part, which describes the time evolution of the CT spin density matrix. 
The dynamics of electron and hole positions are determined by a kinetic Monte Carlo (KMC) algorithm \cite{Voter2007}, whereby the electron and hole migrate via stochastic hops between neighboring molecular sites. 
We restrict the dynamics to include only single particle hops (i.e., electron or hole) and assign hopping rates following the Miller-Abrahams formula~\cite{Miller1960}.
As such, the rate for an electron to hop from site $i$ to site $i'$ while the hole is fixed at site $j$ is given by
\begin{equation}
k_{ij \rightarrow i'j}=\nu_0 \exp \left [ -\frac{(\Delta E_{ij \rightarrow i'j} + \vert \Delta E_{ij \rightarrow i'j} \vert ) }{ 2k_\mathrm{B} T }\right ],
\label{eq:MA}
\end{equation}
where $\nu_0$ is the normalized hopping frequency, $k_\mathrm{B}T$ is the Boltzmann constant times temperature, and $\Delta E_{ij\rightarrow i'j} = E_{i'j}^{(\mathrm{\mathrm{CT}})} - E_{ij}^{(\mathrm{CT})}$.
The hole hopping rate $k_{ij \rightarrow ij'}$ is given by an analogous formula.

The spin dynamics are modeled with an open quantum systems approach in which the electron-hole spin state, $\rho(t)$, is coupled to a bath of harmonic oscillators and propagated via a secular Redfield equation \cite{Blum2012}. 
The spin Hamiltonian in Eq.~\ref{eq:hamiltonian} depends on CT state configuration and therefore the stochastic spatial dynamics of the electron-hole pair imparts a time dependence to the spin Hamiltonian.
In between charge hopping events, however, the electron and hole positions are assumed to be fixed and thus the Hamiltonian of Eq.~\ref{eq:hamiltonian} is static. 
A detailed description of the spin dynamics and the Redfield formalism can be found in the Supporting Information (SI).

The empirical model parameters that define the coarse-grained system can be assigned in a variety of ways.
For instance they can be inferred through the analysis of experimental data or computed via \textit{ab-initio} molecular simulation.
The ability to vary these parameters in order to describe different materials provides the versatility to adapt this model to describe the broad range of systems that exhibit MFEs.
We now demonstrate this versatility by applying our model to investigate a recent set of magnetic field dependent experiments aimed at probing CT-state dynamics.

Recently, Adachi \textit{et al}.\cite{Goushi2012} and Baldo \textit{et al}\cite{Chang2015a, Deotare2015} have developed a donor-acceptor pair of organic dye molecules, 4,4$'$,4$''$-tris[3-methylphenyl(phenyl)amino]-triphenylamine (m-MTDATA) and tris-[3-(3-pyridyl)-mesityl]borane (3TPYMB), which can support electronically excited CT states that can undergo direct singlet radiative recombination.
For thin films blends of these molecules this radiative processes is evident in the photoluminescence (PL), which exhibits a long time ($\sim 30 \mu$s) decay that has been attributed to reverse intersystem crossing from a long lived population of CT triplets. 
Focusing on this long time PL signature, time-resolved fluorescence microscopy has revealed that the PL profile undergoes both a transient spatial broadening and a transient redshift\cite{Deotare2015}, indicating that CT states are mobile along the donor-acceptor interface and sensitive to the presence of static energetic disorder. 
The PL also exhibits pronounced MFEs, indicating that CT state dynamics may involve fluctuations in electron-hole separation. 
These observations led the authors to hypothesize that CT dynamics proceed through the asynchronous motion of localized electrons and holes\cite{Deotare2015}. 
Here we apply our model to this system in order to (\textit{i}) confirm that the hypothesized description of CT state dynamics is consistent with the observed MFEs, and (\textit{ii}) to elaborate on the role of spin dynamics in charge-transfer mediated processes such as photocurrent generation and photoluminescence. 

To adapt our model to this system we utilized a parameterization that was based only on experimentally available data.
The model system included a regular lattice of molecular sites where the lattice spacing was based on the average excluded volume size of the constituent molecules.
As illustrated in Fig.~\ref{fig:schematic}b, the system was divided so that one half of the system contains only donor molecules and the other half contains only acceptor molecules.
We assumed the presence of uncorrelated static energetic disorder, which was represented by assigning values of $E_i^{(\mathrm{LUMO})}$ and $E_j^ {(\mathrm{HOMO})}$ randomly from a Gaussian distribution with standard deviation inferred from spectrocopy.
Experimental data was also used to parameterize the exchange coupling, radiative recombination rate, and the details of spin dynamics. 
A more detailed description of parameter values, their experimental origin, and our treatment of these processes is presented in the Supporting Information (SI).

To simulate CT state PL we generated trajectories that were initiated in a pure singlet state with the electron and hole on adjacent sites at the donor-acceptor interface. 
We generated statistics by sampling many trajectories across many realizations of the static disorder.
Individual trajectories were carried out for a finite observation time ($\tau_\mathrm{obs}=30\mu s$) which was chosen to be approximately the experimental time window in Ref.~\citenum{Deotare2015}, however trajectories could also be terminated at earlier times via a radiative recombination event.
We modeled radiative recombination as a stochastic event with a rate that was proportional to the singlet population and was only allowed if the electron and hole occupied adjacent interfacial sites (see SI for more details). 
 
We simulated transient PL by analyzing the energies and positions of the ensemble of CT states that underwent radiative recombination.  
We find that our model is capable of reproducing the experimentally obtained transient PL data (i.e., spatial broadening and redshift) with near perfect agreement.
A direct comparison of our simulation data to these experimental results can be found in the SI.
Here we narrow our discussion to focus on the unique capability of this model to reveal the effect of applied magnetic field on CT state dynamics.

In the results presented in Ref.~\citenum{Deotare2015} MFEs were quantified in terms of the field dependence of the integrated PL and photocurrent.
We compute integrated PL by first generating an ensemble of trajectories at a given value of $B= \vert \vec{B}^\mathrm{(app)} \vert$ and then evaluating the fraction of trajectories that terminate due to radiative recombination.
Similarly, we relate integrated photocurrent to internal quantum efficiency (IQE) which is evaluated by computing the fraction of trajectories for which the electron-hole separation at $t=\tau_\mathrm{obs}$ exceeds the Coulomb radius (i.e., the distance at which the electrostatic electron-hole interaction is equal to the thermal energy, $k_\mathrm{B}T$). 
Since our model does not include non-radiative loss mechanisms we expect our simulated values to be overestimated relative to experiment.
We have accounted for these unknown loss mechanisms by scaling our results by a field-independent constant.

\begin{figure}[t]   \center
    \includegraphics[width=3.5in]{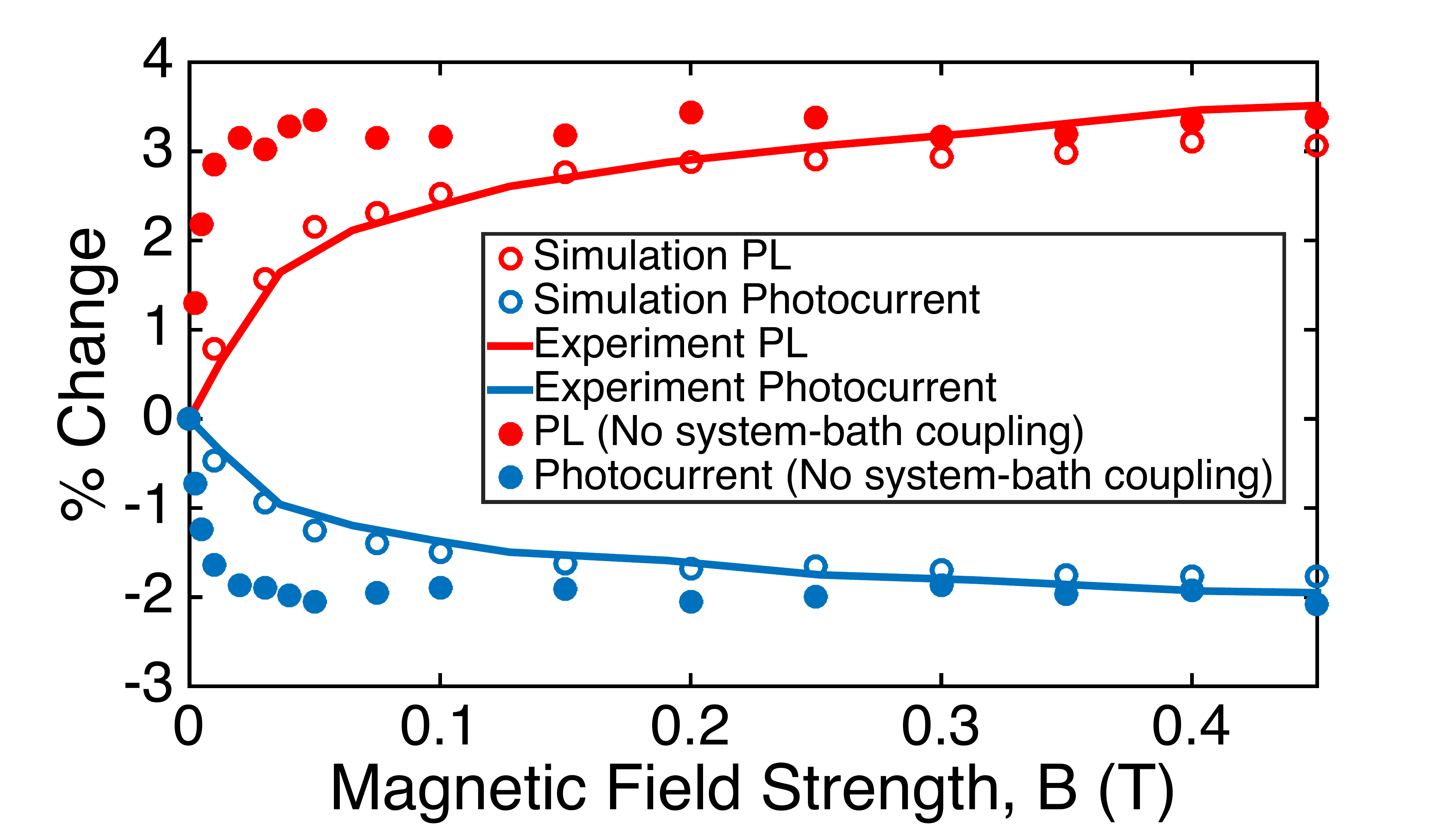}
         \caption{ The magnetic field dependence of photoluminescence (PL) and photocurrent as determined experimently (solid lines) and simulated with our model (unfilled circles). 
         The quantity plotted against the $y$-axis is the percentage change, measured relative to the case where $B=0$.  
         Filled circles correspond to simulated results in the absence of system-bath coupling for spin dynamics. 
          }
    \label{fig:MFE}
\end{figure}

Fig.~\ref{fig:MFE} contains a plot of the percent change in integrated PL and photocurrent as measured experimentally (solid lines) and as predicted from our simulation data (open circles).
Experiments yield an increase in PL with the application of a magnetic field that saturates at fields approaching 0.5T.
There is a corresponding decrease in the integrated photocurrent (more fluorescing CT states leaves fewer free charge carriers for photocurrent generation). 
The simulated CT dynamics accurately reproduce the shape of the experimentally measured MFEs in both the integrated PL and the photocurrent.
The ability of our model to reproduce both the experimentally obtained transient PL and MFEs indicates that our theoretical framework accurately captures the basic physics associated with CT state dynamics in this system.
Building up on this validation we now turn our attention to the ability of this model to reveal information about CT dynamics that are experimentally unavailable. 

To begin we consider the physical origins of singlet-triplet population transfer.
For CT states in systems composed of light molecules (e.g., in the absence of spin-orbit coupling) it is often assumed that this intersystem crossing is driven only by the hyperfine coupling~\cite{Schulten1978,Kersten2011,Kersten2011a,Manolopoulos2013,Lewis2014}.
However, our simulation results reveal that there are alternative spin relaxation pathways that play a significant role in facilitating spin mixing dynamics. 
These spin relaxation pathways are described implicitly in our model in terms of a system-bath coupling in the Redfield relaxation tensor.
This coupling drives the so-called ``spin-flip'' transition, mediating population transfer specifically between the singlet and the $T_0$ triplet state~\cite{Steiner1989}.
If we silence this coupling, then the relatively weak hyperfine field ($\sim 1$mT) is easily overcome by an externally applied field, leading to MFEs that saturate at very small fields.
This is illustrated (filled circles) in Fig.~\ref{fig:MFE}, where the absence of this system-bath coupling results in MFEs that rise sharply and saturate at around $B=10-20$mT, in qualitative disagreement with experimental observations.

\begin{figure}[t]   \center
     \includegraphics[width=3.5in]{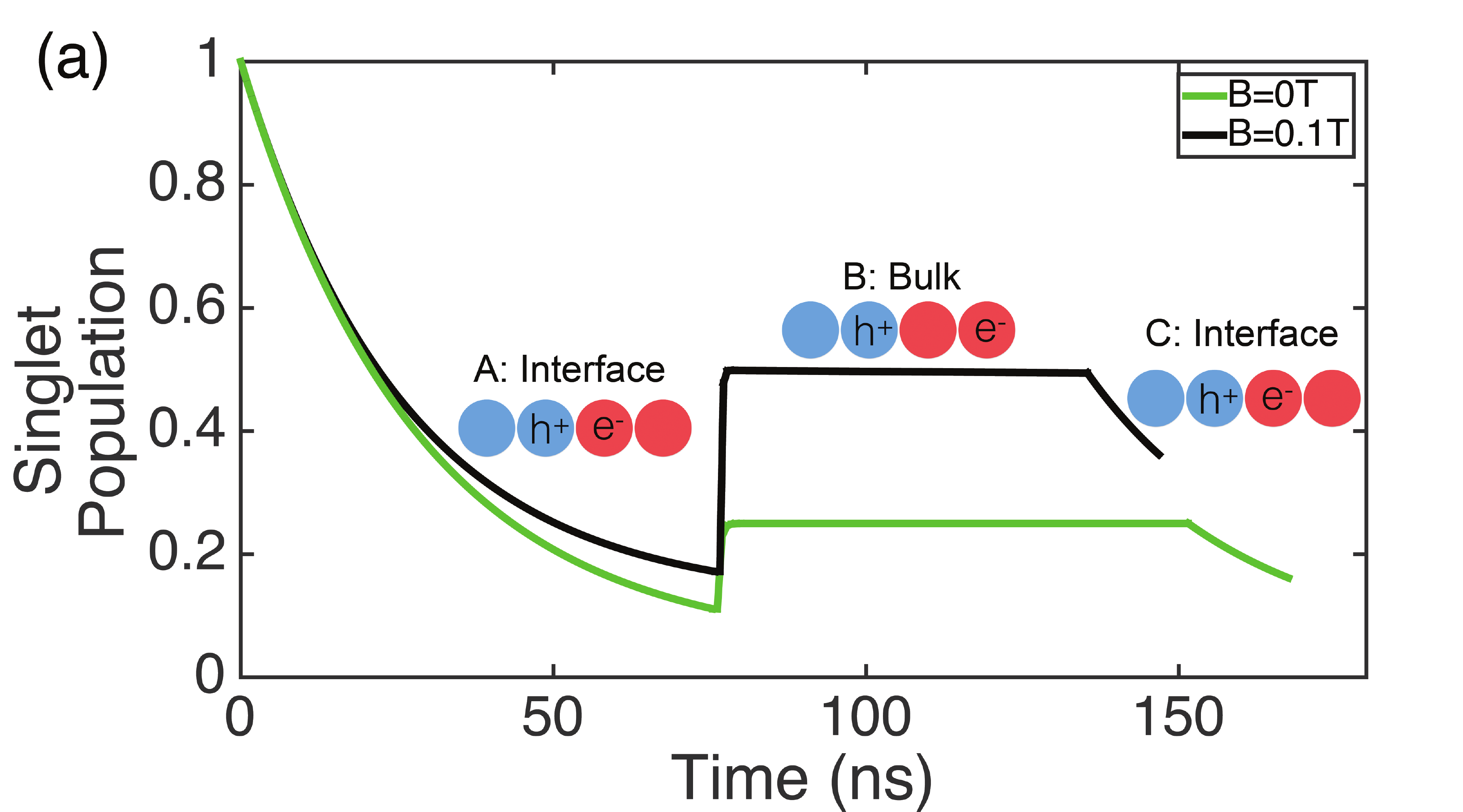}
    \includegraphics[width=3.5in]{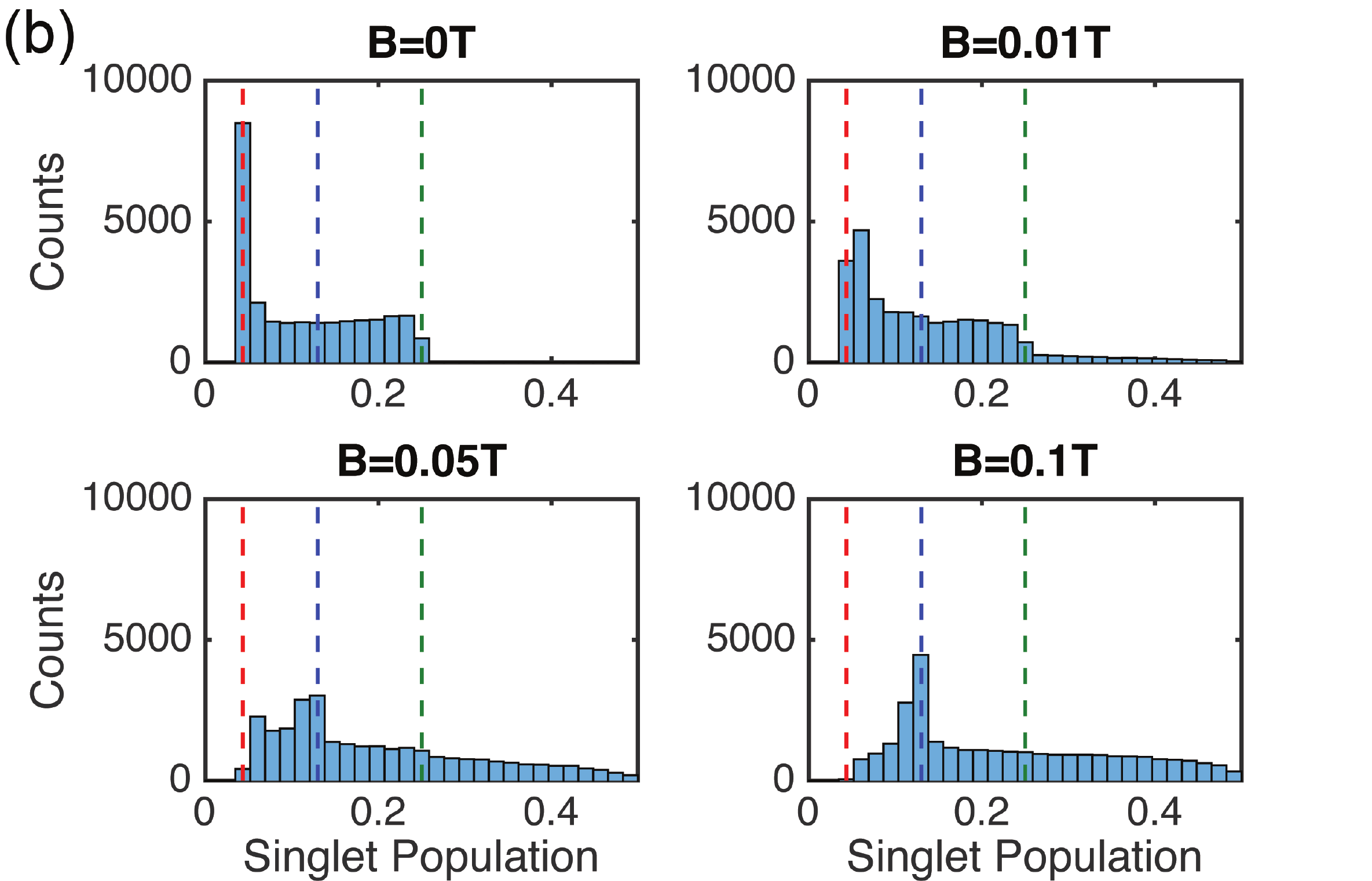}
     \caption{(a) The time-dependence of the singlet population for two typical trajectories, selected to exhibit similar spatial dynamics, carried out under different values of $B=\vert \vec{B}^{\mathrm{app}} \vert$. 
     The inset illustrates the hopping behavior of these trajectories, which each ends abruptly with a radiative recombination event.
     (b) The distribution of the values of singlet population, $\rho_\mathrm{S}$, at the recombination time for trajectories generated at different values of applied magnetic field. 
     The red- and green-dashed vertical lines represent the equilibrium values for the bound ($\langle \rho_\mathrm{S} \rangle=0.05$) and unbound ($\langle \rho_\mathrm{S} \rangle=0.25$) CT states respectively.
     The blue-dashed line represents the value of $\rho_\mathrm{S}$ for the bound CT state in two-state quasi-equilibrium.
     }
      \label{fig:trajectory}
\end{figure}


Using our model we can explore the microscopic fluctuations that give rise to MFEs. 
To illustrate this we consider two representative trajectories each generated at different values of $B$, but exhibiting similar spatial dynamics.
As illustrated in Fig.~\ref{fig:trajectory}a, the trajectories include three distinct segments:
First, in segment A, the electron-hole pair is initiated as singlet state on neighboring sites along the interface.
Next, in segment B, the electron hops away from the interface to form an unbound CT state with a concomitant reduction in the exchange coupling.
Finally, in segment C, the electron and hole reunite on neighboring interfacial sites prior to undergoing radiative recombination.
Although the spatial dynamics of these two trajectories are similar, due to the differing applied magnetic field their spin dynamics differ significantly.
In order to appreciate these differences we consider each trajectory separately, starting with the $B=0$ case (green line in Fig.~\ref{fig:trajectory}a).	

The trajectory is initialized as a bound CT state with an interfacial exchange splitting of $50$meV that lowers the triplet state energy relative to that of the singlet state. 
This energy difference favors the formation of triplet states, with a Boltzmann-weighted singlet density of $\langle \rho_S \rangle \approx 0.05$.
The evolution of the spin state from the initial singlet state is mediated primarily by the system-bath coupling, with a characteristic relaxation timescale of approximately $40$ns. 
Before the spin state can fully relax, however, the system enters segment B by hopping into an unbound CT state configuration.
In our model any unbound state is free of exchange coupling and thus the singlet and triplet states are degenerate.
The associated equilibrium singlet density for the unbound state is $\langle \rho_S \rangle = 0.25$. 
The spin relaxation for this degenerate unbound state is ultrafast, as evident in the rapid equilibration of the singlet population in Fig.~\ref{fig:trajectory}a.
In segment C the CT state re-enters the bound state and proceeds again toward the bound singlet density of $\langle \rho_S \rangle \approx 0.05$.
During this equilibration the CT state undergoes a radiative recombination event, signaling the termination of the trajectory.

For the trajectory generated with $B=0.1$T (black line in Fig.~\ref{fig:trajectory}a) the effect of CT configuration on the equilibrium $\langle \rho_S \rangle$ is identical.
The ability of the spin state to respond to changes in configuration, however, is significantly affected by the presence of the applied magnetic field.
At $B=0.1$T the intra-triplet relaxation occurs on timescales much longer than the length of the trajectory.
Due to this separation in timescales the spin dynamics of this trajectory can be understood in terms of a quasi-equilibrium between the S and $T_\mathrm{0}$ states.
Under this two-state quasi-equilibrium the bound state singlet population approaches $\rho_\mathrm{S}=0.13$ and the unbound state approaches $\rho_\mathrm{S}=0.5$. 
The field-induced slowing of intra-triplet spin relaxation therefore has the effect of both prolonging the redistribution of initial singlet population and, perhaps more importantly, of amplifying the effect of fluctuations in electron-hole separation on the transient singlet population.

The qualitative insight generated by analyzing individual trajectories can be further supported through the statistical analysis of many trajectories.
Fig.~\ref{fig:trajectory}b contains histograms that reveal the distribution of singlet density, $\rho_\mathrm{S}$, amongst the population of fluorescing CT states. 
Each of the four histograms depicted in Fig.~\ref{fig:trajectory}b was generated under different values of $B$.
For the case of $B=0$, the distribution is peaked around $\rho_\mathrm{S}=0.05$ (red dashed line) corresponding to the equilibrium $\langle \rho_S \rangle$ for the bound CT state.
The distribution also includes a tail that extends to $\rho_\mathrm{S}=0.25$ (green dashed line), reflecting the population of CT states that fluoresce shortly after re-entering the bound state, before fully equilibrating. 
This shows that even in the absence of an applied magnetic field a significant portion of luminescent CT excitons exhibit non-equilibrium spin statistics that result directly from fluctuations in electron-hole separation.
As $B$ increases the shape of the histograms change to reflect two-state (S and $T_0$) quasi equilibrium, with peak at  $\rho_\mathrm{S}=0.13$ (blue dashed line), that results from field-induced slowing of spin mixing dynamics. 
The field-dependence of these histograms highlight the microscopic origin of observed MFEs, namely that field-induced non-equilibrium spin statistics serve to enhance the singlet population and thereby the PL yields.


\begin{figure}[t]   \center
    \includegraphics[width=3.5in]{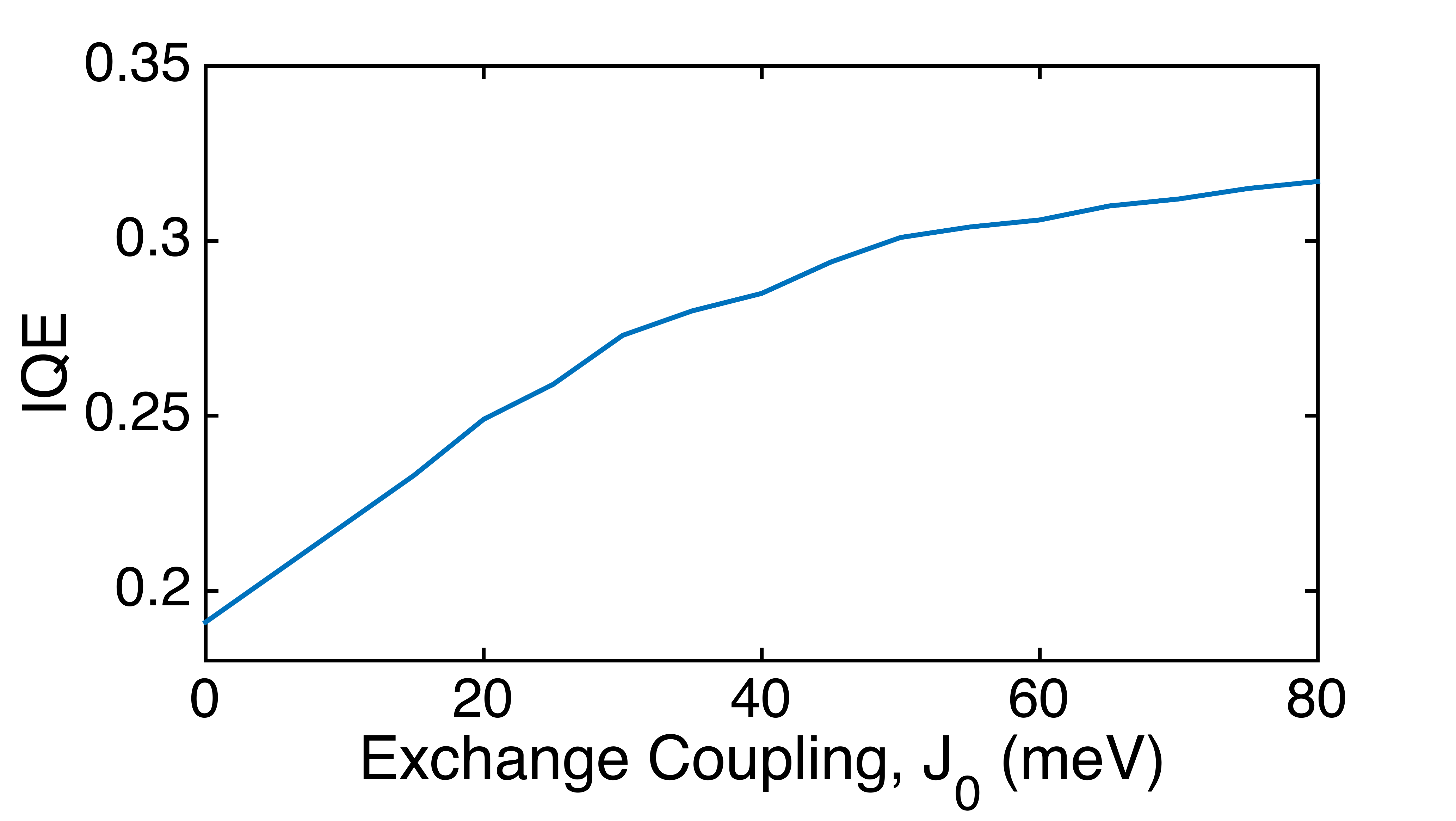}
     \caption{The dependence of photovoltaic IQE on the magnitude of the interfacial exchange splitting as predicted from simulations carried out on our model system. }
          \label{fig:IQE_exchange}
\end{figure}


The analysis described above clearly demonstrates the importance of the time dependent exchange splitting on the spin dynamics. 
With our model we can evaluate the role of this interfacial exchange splitting, $J_0$, on predicted device performance. 
To do this we have carried out a series of simulations each with varying values of $J_0$.
Our findings, shown in Fig.~\ref{fig:IQE_exchange}, illustrate that larger $J_0$ is beneficial for OPV performance.
Specifically, as $J_0$ increases from 0meV to 80meV,  the simulated IQE increases by nearly 70$\%$. 
Qualitatively, this efficiency increase arises because energetically favorable triplet states are spin protected from radiative recombination and thus the electron and hole have more time to diffuse away from each other to generate free charges. 
By initiating CT states as spin-equilibrated free charges at the simulation boundary our model can be utilized to simulate electroluminescence. 
Increasing exchange coupling was found to reduce electroluminescence efficiency, which is consistent with experimental observations reported in Ref.~\citenum{Zhang2014}. 

The model presented here offers an efficient and versatile tool that can be used to relate difficult to interpret magnetic field sensitive experiments to the microscopic fluctuations of excited electron-hole pairs.
By applying this model to the donor-acceptor blend described in Ref.~\citenum{Deotare2015} we have highlighted how MFEs emerge from the details of spin mixing dynamics. 
Furthermore, we have illustrated how the interplay between spin and spatial dynamics contribute to CT state dynamics and experimentally observed MFEs.
The insight we have drawn highlights the benefit of simple models in guiding our intuition around complex physical systems.
This model can be applied in a straightforward manner to describe the optoelectronic properties of other CT-mediated processes, perhaps those that involve more complicated interfacial molecular morphology. 

\textit{Acknowledgements} -- The authors would like to thank Eric Hontz, Wendi Chang, Parag Deotare, Daniel Congreve, Marc Baldo and Troy Van Voorhis for helpful discussions. This work was supported by startup funds from the Department of Chemistry at Massachusetts Institute of Technology.

\bibliography{spin_references}

\newpage
\begin{figure}[t]   \center
    \includegraphics[width=3.5in]{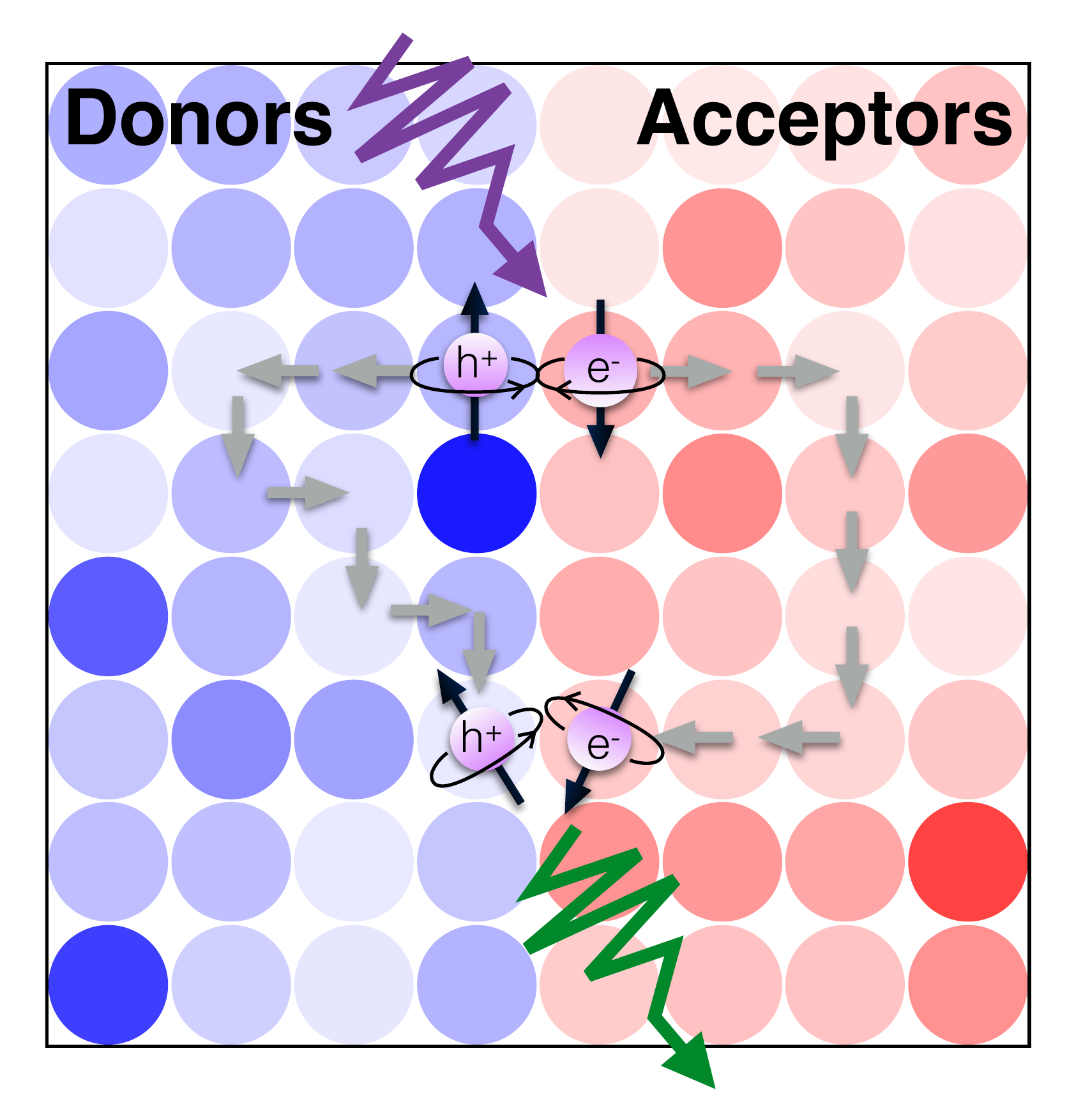}
         \caption{ Table of Contents graphic.}
    \label{fig:TOC}
\end{figure}

\end{document}


\topmargin -13mm 
\title{Magnetic Field Effects in Organic Semiconductors} 
\author{Chee Kong Lee}
\affiliation{Department of Chemistry, Massachusetts
Institute of Technology, Cambridge, Massachusetts 02139, USA}
\author{Liang Shi} 
\affiliation{Department of Chemistry, Massachusetts
Institute of Technology, Cambridge, Massachusetts 02139, USA}
\author{Adam Willard} \email{awillard@mit.edu}
\affiliation{Department of Chemistry, Massachusetts
Institute of Technology, Cambridge, Massachusetts 02139, USA}

\begin{abstract} 
xxx
\end{abstract}

%


\appendix
\section*{Supporting Information}
\subsection{Spatial Dependence of Exchange Splitting and Radiative Recombination }
We model the exchange coupling as a step function with the form,
\begin{equation}
  	J(\vert \vec{r}_i - \vec{r}_j\vert ) = \left\{ \begin{array}{ll}
		J_0, & \text{if $i$ and $j$ are nearest neighbors (i.e., the bound CT state),} \\
          	0, & \text{otherwise (i.e., the unbound CT state)}. \end{array} \right.
\end{equation}
We estimate $J_0=50$meV based on the experimental data \cite{Goushi2012, Deotare2015}.
The standard kinetic Monte Carlo (KMC) algorithm was modified to include the ability for the electron and hole to radiatively recombine.
This recombination can only occur when the electron and hole reside on neighboring interfacial molecules, and the probability of recombination is proportional to
the singlet density.
Formally, this conditional relaxation pathway can be expressed in terms of a KMC processes with a rate,
\begin{equation}
  	k_{ij}^{(\mathrm{PL})} = \left\{ \begin{array}{ll}
		k_\mathrm{PL}, & \text{if $i$ and $j$ are nearest neighbors,} \\
          	0, & \text{otherwise}, \end{array} \right.
\end{equation}
where the dependence of the photoluminescence rate on spin state is described by treating $k_\mathrm{PL}$ as a stochastic random variable with the properties,
\begin{equation}
  	k_\mathrm{PL} = \left\{ \begin{array}{ll}
		\tau_\mathrm{PL}^{-1}, & \text{with probability $\rho_\mathrm{S}$},\\
          	0, & \text{with probability $1-\rho_\mathrm{S}$}. \end{array} \right.
\end{equation}	
Here $\rho_\mathrm{S}$ represents the singlet projection of the two-spin quantum density matrix, $\rho$.
The KMC parameters, $\nu_0=15\mu \mathrm{s}^{-1}$ and $\tau_\mathrm{PL}=8 \mu \mathrm{s}^{-1}$, where chosen based on transient photoluminescence measurements \cite{Deotare2015}. 

Our lattice model comprises 300 donor sites and 300 acceptor sites, divided by a linear donor-acceptor interface of 
20 interfacial donor-acceptor pairs. 
The lattice spacing is chosen to be 2.5 nm, roughly approximating the excluded-volume diameter of the molecules in the experiment. 
We use $E^{\rm (HOMO)}_j=5.1eV$ and $E^{\rm (LUMO)}_j=3.3eV$ for the HOMO energy of donor molecules and LUMO energy of acceptor molecules, respectively. 
Both energies are assigned Gaussian disorders with standard deviation of $60 \mathrm{meV}$ to describe the inhomogeneous broadening. 

\subsection{Quantum Master Equation}
We use the standard system-plus-bath approach to describe the spin relaxation. The total system and bath Hamiltonian is given by
\begin{equation}
   H^{\rm(Tot)}=H^{\rm (S)}+H^{\rm (B)}+H^{\rm(SB)},
\end{equation} 
where the three terms represent the Hamiltonians of the system, the bath, and system-bath coupling, respectively.
The system Hamiltonian, $H^{\rm (S)}$, is described by the electron (indexed $e$) and hole (indexed $h$) spin operators and its explicit form is given in Eq. (3) in the 
main text. 
We assume that the electron and hole are independently coupled to its own harmonic bath, thus 
\begin{eqnarray}
H^{\rm{(B)}} &=& \sum_{\alpha=e,h}\sum_{n}\omega_{\alpha}^{(n)} b^{(n)\dagger}_{\alpha}b_{\alpha}^{(n)}, \\
H^{\rm{(SB)}} &=& \sum_{\alpha=e,h} \sum_{n}g_{\alpha}^{(n)}  S_\alpha^{z} (b^{\dagger(n)}_{\alpha} + b_{\alpha}^{(n)}) , 
\end{eqnarray}
where $\omega_{\alpha}^{(n)}$ and $b^{(n)\dagger}_{\alpha}$($b_{\alpha}^{(n)}$) are the frequency 
and the creation (annihilation) operator of the $n$-th mode of the harmonic bath 
coupled to electron or hole with coupling strength $g_{\alpha}^{(n)}$, respectively.
We assume the coupling constants are identical for both the electron and hole, i.e.  $g_{\alpha}^{(n)}=g^{(n)}$. Additionally, we choose a Drude-Lorentz spectral density,
$J(\omega)=\frac{\pi}{2}\sum_n \frac{|g^{(n)}|^2}{\omega_n}\delta(\omega - \omega_n) = 2 \gamma \omega_c \frac{\omega}{\omega^2 +\omega_c^2}$,  
where $\gamma$ is the dissipation strength and $\omega_c$ is the cut-off frequency.

A perturbation approximation can be applied in terms of the system-bath coupling 
leading to a standard Redfield quantum master equation of the reduced density matrix~\cite{Redfield1957, Nitzan2006}:
\begin{eqnarray} 
   \frac{d \rho_{\mu \nu}(t)}{dt} &=&   - \iu \,\omega_{\nu \mu} \rho_{\mu \nu}(t)
   + \sum_{\mu'\nu'}R_{\mu \nu, \mu'\nu'} \rho_{\mu' \nu'}(t), 
\end{eqnarray} 
where the Markovian approximation has been employed. 
The Greek indices denote the eigenstates of the  system Hamiltonian, 
i.e. $H^{\rm (s)} |\mu\rangle = E_\mu |\mu\rangle$ and 
$\omega_{\mu\nu} = (E_\mu - E_\nu)/\hbar $.
The Redfield tensor, $R_{\mu\nu,\mu'\nu'}$, describes the spin relaxation and can be
expressed as
\begin{align} 
   R_{\mu\nu,\mu'\nu'} &=  \Gamma_{\nu'\nu,\mu\mu'} +
   \Gamma_{\mu'\mu,\nu\nu'}^{*}
    -\delta_{\nu\nu'}\sum_{\kappa}
   \Gamma_{\mu\kappa,\kappa\mu'} -\delta_{\mu\mu'} \sum_{\kappa}
   \Gamma_{\nu\kappa,\kappa\nu'}^{*};\\
 \Gamma_{\mu\nu,\mu'\nu'}  &= \sum_{\alpha=e,h}  \langle \mu | S_\alpha^{z} | \nu \rangle \langle \mu' | S_\alpha^{z} | \nu' \rangle
   K(\omega_{\nu'\mu'})
   \;, \label{eq:ptre_rates}
\end{align}
where $K(\omega)$ is the half-Fourier transform of the phonon bath correlation function
\begin{equation} K(\omega) = \int^{\infty}_{0} \frac{d\omega}{\pi} J(\omega) \Big[ \coth \Big( \hbar \beta \omega/2\Big) \cos(\omega t) - i \sin(\omega t)\Big],
\end{equation}
where $\beta =\frac{1}{k_B T}$ is the inverse thermal energy. 
Finally, we invoke the secular approximation, dropping the terms in the Redfield tensor, $R_{\mu\nu,\mu'\nu'} $, for which $\omega_{\mu\nu} -\omega_{\mu'\nu'}\neq 0$.
Employing the secular approximation ensures the positivity of the reduced density matrix, i.e. the diagonal matrix elements are always positive~\cite{Breuer2002,Blum2012}. 
In addition to preserving positivity, the secular approximation also guarantees the long-time equilibrium state is given by the Boltzmann state of the system Hamiltonian, 
$\rho(t \rightarrow \infty) = \frac{\mbox{e}^{-\beta H^{\rm (s)}} }{\mbox{tr}[\mbox{e}^{-\beta H^{\rm (s)}}]}$.
In this letter, we use $\omega_c=0.004$meV and $\gamma=3\times 10^{-8}$meV. The hyperfine magnetic field, $\vec{B}^{\rm (hf)}$, is drawn from a 3D Gaussian distribution with a standard deviation of $1 \mathrm{mT}$.  

\subsection{Time-Dependent Photoluminescence Properties}
\subsubsection{Transient Photoluminescence}
\begin{figure}[H]   \center
    \includegraphics[width=3.5in]{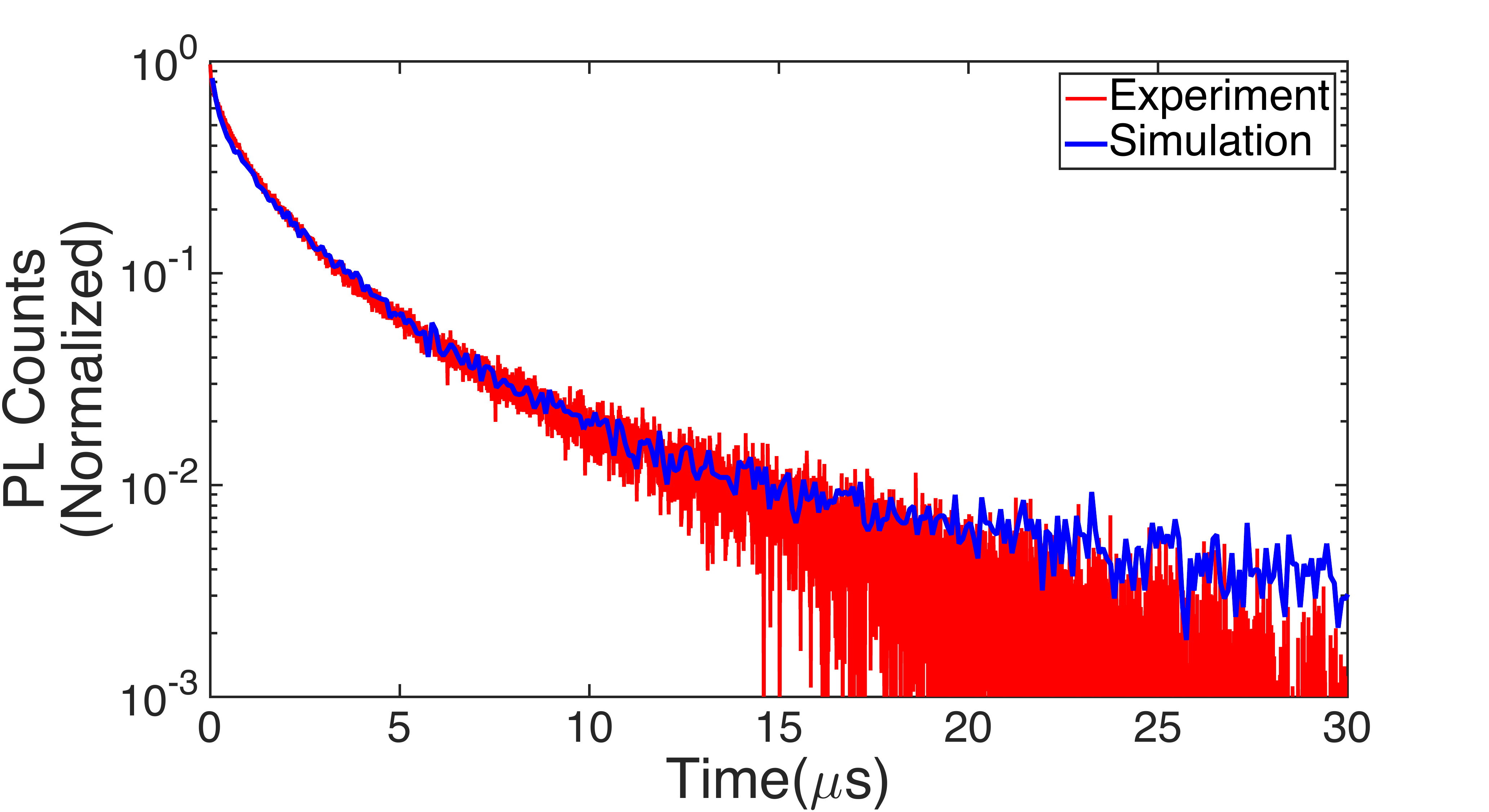} 
         \caption{The simulated(blue) and experimental(red) transient decay of photoluminescence (normalized).}
          \label{fig:transient}
\end{figure}

We compute the transient photoluminescence (PL) by counting the number of recombined CT excitons per unit time in the KMC simulations.
The comparison with experimental data from Ref.~\cite{Deotare2015}, plotted in Fig. \ref{fig:transient}, allows us to estimate the singlet decay rate, $k_{PL}$.

\subsubsection{Mean Squared Displacement}
\begin{figure}[H]   \center
    \includegraphics[width=3.5in]{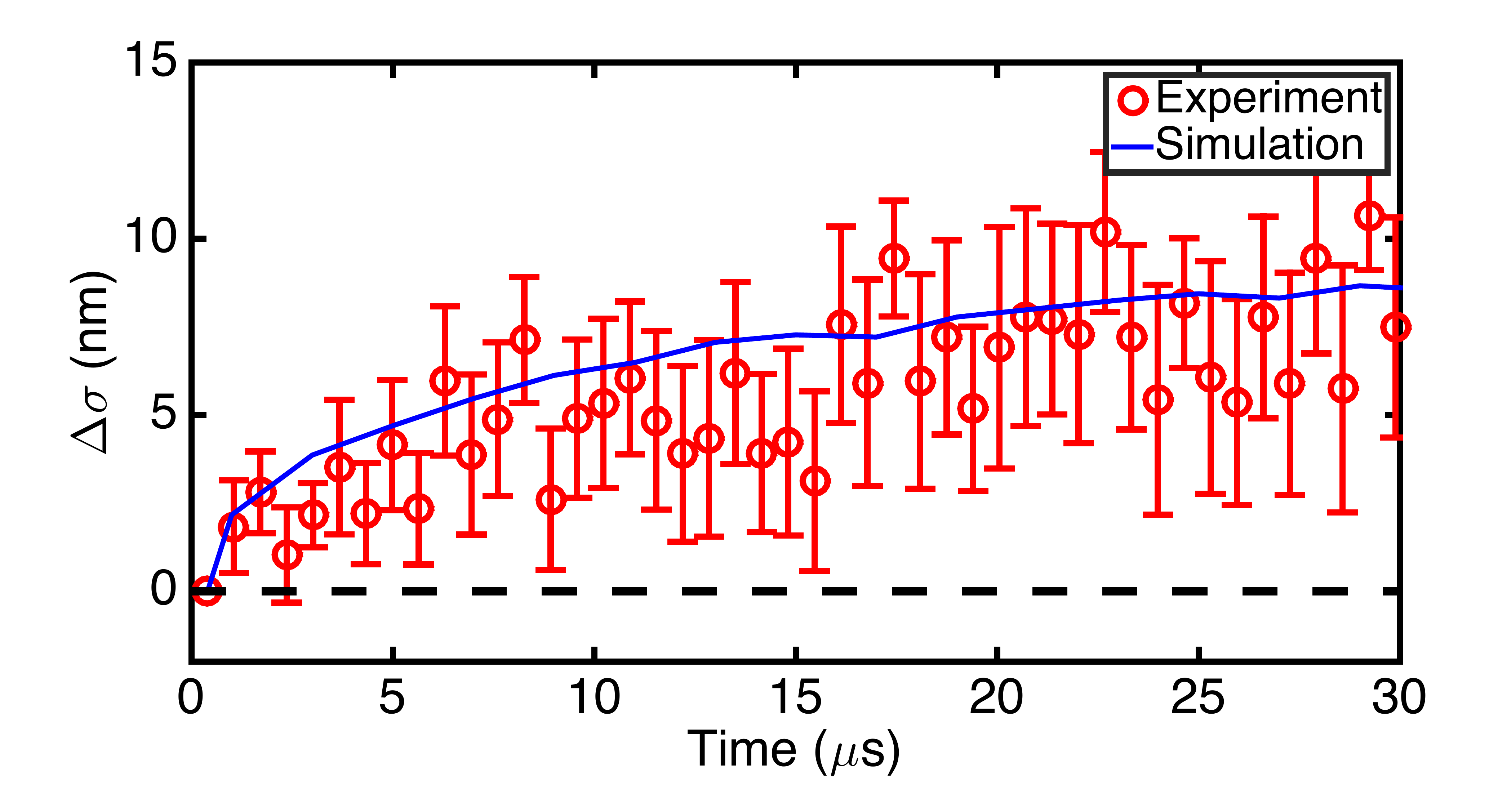}
         \caption{
          The simulated mean squared displacement of CT excitons (blue). 
          The red circles show the experimental values of the standard deviation of spatial broadening Gaussian function of PL. 
          The error bars indicate the standard error between four independent diffusing imaging measurements \cite{Deotare2015}.}
          \label{fig:MSD}
\end{figure}

In our simulations, the displacement of a CT exciton is defined as the distance between the initial CT state and the eventual recombination site along the interface. 
With this definition, the mean squared displacement from the KMC simulations and the experimental data are plotted in Fig.~ \ref{fig:MSD}.
The increase in mean squared displacement indicates that the CT excitons can move geminately over distance of several molecules (5-10nm).

\subsubsection{Transient Redshift}
\begin{figure}[H]   \center
    \includegraphics[width=3.5in]{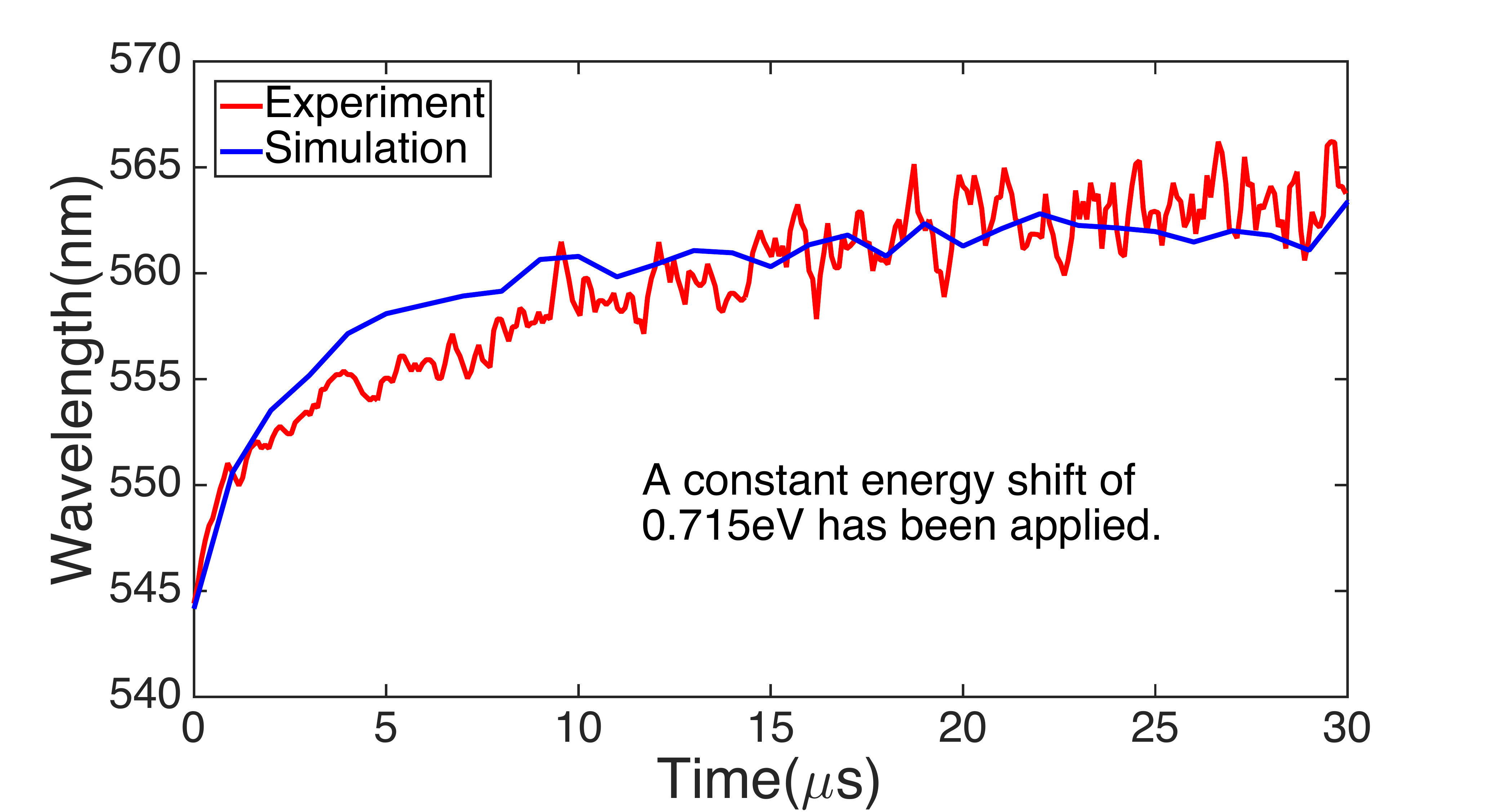}
         \caption{The centroid of the experimental data (red) is compared to the spectral shift predicted by the kinetic Monte Carlo simulation (blue). The transient redshift of the CT exciton emission wavelength indicates the CT excitons travel to lower energy sides. }
          \label{fig:spectral}
\end{figure}

By computing the averaged energy of the recombined CT excitons, see Eq. (1) in main text, our model is capable of reproducing
 the time dependence of the CT exciton emission wavelength. 
The comparison between the simulated and experimental results is plotted in Fig.~\ref{fig:spectral}. 
The redshift of the emission wavelength indicates the preference of the electron and hole to diffuse to lower-energy interfacial sites, 
and therefore a manifestation of nanoscale disorder at the interface.

\bibliographystyle{achemso}
\bibliography{SI_references}